# THERMALLY-TARGETED ADSORPTION AND ENRICHMENT IN MICROSCALE HYDROTHERMAL PORE ENVIRONMENTS

**Aashish Priye, Yassin A. Hassan, and Victor M. Ugaz**
*Department of Chemical Engineering, Texas A&M University, College Station, TX  77843, USA*

**ABSTRACT**
The unique ability of chaotic advection under micro-scale confinement to direct chemical processes along accelerated kinetic pathways has long been recognized. But practical applications have been slow to emerge because optimal results are often counter-intuitively achieved in flows that appear to possess undesirably high disorder. Here we demonstrate how thermally actuated chaotic phenomena within these microenvironments are capable of establishing a continuous conveyor transporting chemical compounds from the bulk to targeted locations on solid boundaries where they become greatly enriched. These findings intriguingly suggest that microscale chaotic advection may offer a new mechanism to explain emergence of biomolecular complexity from dilute organic precursors in the prebiotic milieu—a key unanswered question in the origin of life. We further show how these flows can be rationally designed and harnessed to execute bulk biochemical processes with a level of robustness previously thought unattainable.

**INTRODUCTION**
The "RNA world" theory offers a widely accepted framework to explain spontaneous emergence of biochemical complexity from elementary building blocks likely to have been present under prebiotic conditions. But attainment of the high-level structure and function associated with RNA and DNA would have required that these organic precursors become sufficiently enriched to initiate condensation polymerization. This poses a conundrum because concentrations of these compounds in the prebiotic ocean were incredibly dilute, strongly favoring hydrolytic decomposition over polymeization. Mineral and clay surfaces may have played an vital role in overcoming these limitations by providing a synergistic combination of enhanced concentration via surface adsorption and catalytic activity capable of directing rections along specific pathways. This view is supported by evidence that montmorillonite rocks are able to catalyze synthesis of both polypeptides and RNA oligomers from organic precursors.

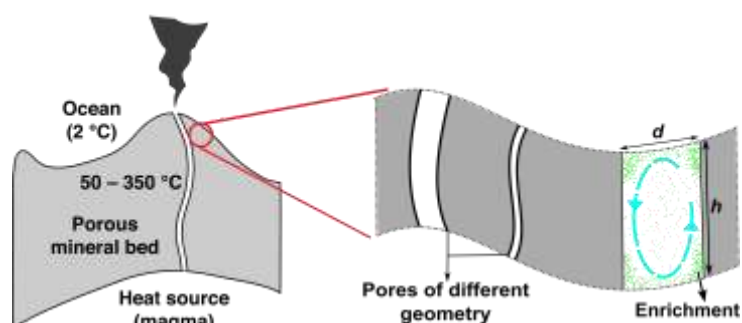

*Figure 1. Microscale chaotic advection provides a strong driving force for thermally actuated transport and reactions. Rock formations near hydrothermal vents lining the ocean floor contain embedded pore networks with microenvironments that impose thermal and geometric conditions robustly capable of sustaining internal convective flow fields. These flows display a rich spectrum of 3D chaotic trajectories that continually shuttle chemical species from the bulk fluid to targeted sites on the pore surfaces where they experience accelerated adsorption and enrichment.*

**THEORY**
These mineral formations are naturally found near hydrothermal vents where geothermally heated water (50 – 350 °C) erupts through the sea floor into cold (2 °C) oceanic surroundings, and contain intricate embedded pore networks with characteristic length scales ranging from μm to cm at aspect ratios (height ($h$) / diameter ($d$)) of 1 – 1,000. The steep temperature gradients (1 – 50 K/mm) in such environments provide a strong driving force to initiate and sustain convecive flow. These uniquely favorable attributes have motivated previous efforts to explore the possibility that hydrothermal pore networks could function as molecular traps capable of concentrating molecules via the coupled action of laminar (2D) thermal convection and thermophoresis, albeit in relatively small geometries (diameter ~ 100 μm) [1]. More recent work, however, has revealed emergence of unexpectedly complex flow fields that are not captured by the laminar 2D picture conventionally assumed.[2-9] These 3D chaotic advection phenomena act over a much broader range of length scales extending beyond the thermophoretic regime, raising the intriguing possibility that they could have functioned in a previously unappreciated way as a highly efficient conveyor for targeted surface enrichment of organic precursors (Fig. 1).

**EXPERIMENTAL**

Adsorption studies were performed using pore-mimicking transparent cylindrical acrylic cells (1.5 mm dia.) mounted in an apparatus that permitted the upper and lower surface temperatures to be independently controlled to impose a vertical gradient. The inner walls of each cell were coated with bovine serum albumin by first sealing the lower surface using thin aluminum tape and rinsing the interior with water. A 10 mg/ml aqueous BSA solution was then loaded, incubated for 5 min, and removed. Coating stability was verified using FITC-BSA. A 10x aqueous dilution of 1 μm dia. carboxylate-modified polystyrene microspheres was then pipetted into the cell and the top surface was sealed with aluminum tape. The filled cells were loaded into the convective apparatus after preheating the upper and lower surfaces to desire temperatures (top 55 °C; bottom 95 °C) followed by 15 minutes of convective flow. After cooling the cell, the sealing tape was removed and the remaining liquid was dried by placing the cells in a 50 °C oven for 10 min. Adsorption profiles were imaged using an Olympus SZX-12 fluorescence microscope with GFP filter set, and the corresponding intensity data were extracted using ImageJ software.

A computational fluid dynamics model was formulated in 3D to simulate Rayleigh Bénard convection in microscale cylindrical reactors to resolve the complex convective flow fields in a series of pore geometries for an imposed vertical temperature difference of 50 °C. An ensemble of 300 randomly distributed Lagrangian passive tracers were simulated in 3D during 5 min of flow, and the location where each trajectory penetrated a 50 μm adsorption boundary layer was recorded to obtain surface enrichment profiles. The anisotropy associated with each profile was quantified in terms of a focusing fraction $f$ expressing the relative amount of surface adsorption localized near the upper and lower boundaries, enabling a parametric map to be constructed depicting the extent of targeted adsorption achievable across a broad range of pore geometries. Adsorption rates at the pore sidewalls was captured by coupling the flow equations with a liquid phase first-order adsorption kinetic model representative of chemical synthesis on mineral adsorbates [10]. The adsorbing species was assumed to be dispersed in water at $10^{-7}$ M.

**RESULTS AND DISCUSSION**

We begin by quantifying the interplay between chaotic convective transport and surface adsorption in a pore-mimicking cylindrical Rayleigh Bénard system across a broad range of size scales representative of the hydrothermal environment (Fig. 2). Under these conditions, the accessible states of fluid motion can be mapped in terms of the aspect ratio ($h/d$) and the dimensionless Rayleigh number ($Ra = [g\ \beta\ (T_2 - T_1)\ h^3] / \nu\ \alpha$, where $\beta$ is the fluid's thermal expansion coefficient, $g$ is gravitational acceleration, $T_1$ and $T_2$ are the temperatures of the top (cold) and bottom (hot) surfaces respectively, $h$ is the height of the fluid layer, $\alpha$ is the thermal diffusivity, and $\nu$ is the kinematic viscosity).

Our simulations reveal that highly focused enrichment (i.e., distribution profiles characterized by distinct bands near the upper and lower pore boundaries) is achievable over a wide spectrum of conditions ranging from laminar to chaotic advection (indicated by emergence of complex flow trajectories).[2] These bimodal adsorption profiles become progressively distorted under conditions where $h/d$ and $Ra$ are simultaneously large, although preferential accumulation near the upper and lower pore boundaries is generally retained. The blue colored region near the bottom right corner of the map in Fig. 2a indicates conditions where the thermal driving force is insufficient to initiate convective motion (i.e., molecular diffusion is the sole transport mechanism). We experimentally verified the simulation predictions by coating the inner walls of cylindrical pore-mimicking flow chambers with bovine serum albumin (BSA). Not only did we observe nearly complete transport of the dispersed microspheres to the sidewalls upon application of a convective flow, the resulting surface fluorescence profiles display targeted enrichment within discrete bands near the top and bottom of the chamber (Fig. 2). Taken together, these results suggest a robust driving force for focused surface adsorption under hydrothermally relevant conditions.

While the results in Fig. 2a quantify the achievable steady-state adsorption profiles, they do not convey information about the rate of transport to the pore surface. We expanded our 3D flow simulations to incorporate a coupled kinetic model that enabled us to track the time-resolved surface concentration of adsorbed species. These results closely resemble the response to a step input of first order; $C / C_{max} = (1 - \exp(t / \tau))$, where $C$ and $C_{max}$ are the instantaneous and final equilibrium (i.e., corresponding to saturation of active sites) surface concentrations respectively, $t$ is time, and $\tau$ is a constant representing the time to reach 63.2% of the asymptotic value. We also determined time resolved surface concentrations associated with purely diffusive transport by analytically solving the transient 2D diffusion equation in cylindrical coordinates. Time constants associated with convection (flow; $\tau_{conv}$) and diffusion (no flow; $\tau_{diff}$) could then be extracted from these data. We evaluated our kinetic model across the same ensemble of flow conditions employed in our analysis of steady-state adsorption to generate the parametric map shown in Fig. 2b. These results indicate that surface adsorption is greatly accelerated in large diameter pores where the flow field is most likely to be characterized by chaotic advection (i.e., in the upper left quadrant of Fig. 2b).[3] At first glance, this finding appears to be at odds with Fig. 2a where the most pronounced focusing near the top and bottom of the pores is observed under predominantly laminar conditions near the onset of convective flow (i.e., immediately adjacent to the blue "no-flow" regime in the lower right corner of Fig. 2a). The accelerated kinetics we observe in more disordered flows likely reflects an additional mixing effect attributable to the onset of chaotic advection that more efficiently transports fluid within the entire pore volume toward the bounding surfaces.[2] But further comparison of Figs. 2a and 2b reveals that pores displaying the fastest adsorption kinetics (i.e., $1 < h/d < 4$, $h > 5$ mm in Fig. 2b) simultaneously generate strong targeted enrichment, suggesting an extensive parameter space where enrichment and adsorption kinetics are synergistically enhanced.

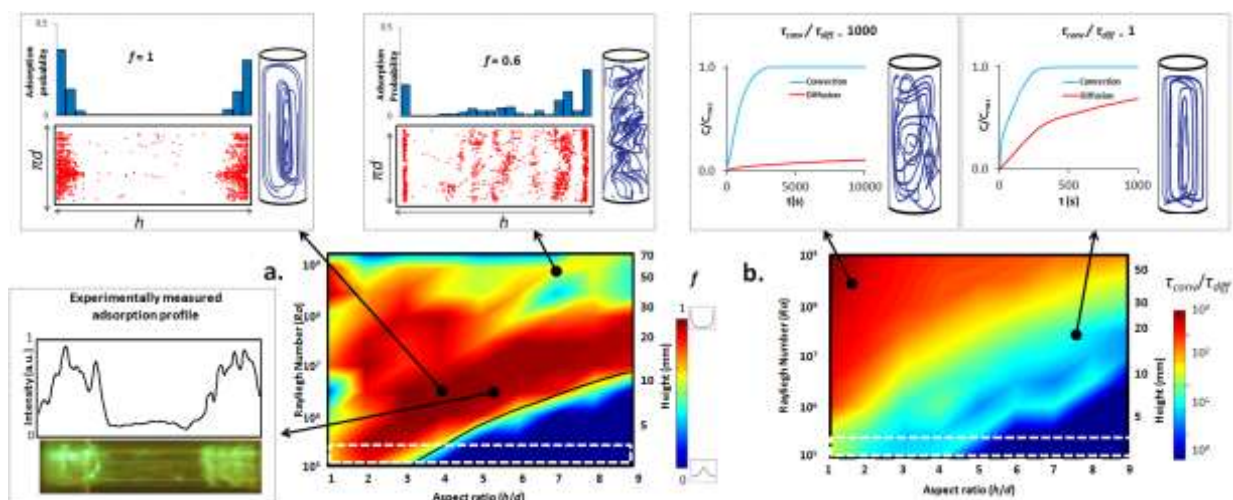

*Figure 2. (a.) Targeted enrichment: Computational simulations enable the extent of targeted enrichment, quantified in terms of a focusing fraction f, to be parametrically plotted in terms of the Rayleigh number and aspect ratio. Representative flow trajectories and surface adsorption are shown above the parametric plot corresponding to values of f ranging from 1 (anisotropic adsorption localized near the upper and lower surfaces) and 0.6. An experimentally obtained adsorption profile (left) reveals localized accumulation of fluorescent carboxylated microspheres near the upper and lower surfaces of the cell, in agreement with the simulations. (b.) Accelerated surface adsorption kinetic: Computational simulations incorporating a kinetic surface adsorption model enable the time-resolved surface accumulation of chemical species to be quantified under conditions relevant to the prebiotic environment. A parametric plot reveals that adsorption kinetics can be enhanced by up to 1000 fold in the presence of a convective transport at the onset of chaotic advection in the corresponding flow trajectories. Pore geometries associated with the flow trajectories are not depicted to scale in order to facilitate comparison between them. Conditions relevant to thermophretic trapping, are denoted at the bottom of the parametric plots (white dashed box).*

## CONCLUSION

Our findings support the hypothesis that chaotic thermal convection provides a robust driving force for enrichment of prebiotic precursors at catalytically active mineral surfaces within hydrothermal pores. We highlight some key features that distinguish the chaotic convective transport mechanism introduced here from thermophoretic trapping[1]. First, we remark that cavities formed within the cracks of mineral deposits and voids encountered between pillow lavas display characteristic size scales that are closely aligned with the convective regime (thermophoretic effects predominate in ~ 100 μm diameter pores,[1] indicated by the narrow dashed white region at the bottom of the parametric plots in Figs. 2a and 2b). Convective transport therefore significantly broadens the range of geometries capable of supporting chemical synthesis to fully encompass realistic pore size distributions encountered in hydrothermally relevant mineral formations. The size scales and boundary conditions associated with thermophoretic trapping may therefore make it best suited to describe phenomena occurring near the surfaces of hydrothermal mineral formations, whereas chaotic thermal convection may dominate throughout the bulk interior spaces. Consequently, these dual mechanisms may function in tandem to synergistically enhance enrichment of chemical precursors pertinent to prebiotic synthesis.

## ACKNOWLEDGEMENTS
This work was supported in part by the US National Science Foundation under grant CBET-1034002.